\documentclass[aps, reprint,amsmath,amssymb,superscriptaddress]{revtex4-1}
\usepackage{graphicx}
\usepackage{dcolumn}
\usepackage{bm}
\usepackage{amsmath}

\begin{document}

\title{Density of states deduced from ESR measurements on low-dimensional nanostructures; benchmarks to identify the ESR signals of graphene and SWCNTs}

\author{P\'{e}ter Szirmai}
\affiliation{Budapest University of Technology and Economics, Institute
of Physics and Condensed Matter Research Group of the Hungarian Academy of
Sciences, H-1521 Budapest, Hungary}
\author{G\'{a}bor F\'{a}bi\'{a}n}
\affiliation{Budapest University of Technology and Economics, Institute
of Physics and Condensed Matter Research Group of the Hungarian Academy of
Sciences, H-1521 Budapest, Hungary}
\author{Bal\'{a}zs D\'{o}ra}
\affiliation{Budapest University of Technology and Economics, Institute
of Physics and Condensed Matter Research Group of the Hungarian Academy of
Sciences, H-1521 Budapest, Hungary}
\author{J\'{a}nos Koltai}
\affiliation{Department of Biological Physics, E\"{o}tv\"{o}s
University, P\'{a}zm\'{a}ny P\'{e}ter s\'{e}t\'{a}ny 1/A, H-1117
Budapest, Hungary}
\author{Viktor Z\'{o}lyomi}
\affiliation{Research Institute for Solid State Physics and Optics, Hungarian Academy
of Sciences, P.O. Box 49, H-1525 Budapest, Hungary}
\affiliation{Department of Physics, Lancaster University, Lancaster,
LA1 4YB, United Kingdom}
\author{Jen\H{o} K\"{u}rti}
\affiliation{Department of Biological Physics, E\"{o}tv\"{o}s
University, P\'{a}zm\'{a}ny P\'{e}ter s\'{e}t\'{a}ny 1/A, H-1117
Budapest, Hungary}
\author{Norbert M. Nemes}
\affiliation{GFMC, Departamento de F\'{\i}sica Aplicada III, Univesidad Complutense, 28040
Madrid, Spain}
\author{L\'{a}szl\'{o} Forr\'{o}}
\affiliation{Laboratory of Physics of Complex Matter, FBS Swiss Federal Institute of Technology (EPFL), 1015 Lausanne, Switzerland}
\author{Ferenc Simon}
\email{ferenc.simon@univie.ac.at}
\affiliation{Budapest University of Technology and Economics, Institute
of Physics and Condensed Matter Research Group of the Hungarian Academy of
Sciences, H-1521 Budapest, Hungary}

\begin{abstract}
Electron spin resonance (ESR) spectroscopy is an important tool to characterize the ground state of conduction electrons and to measure their spin-relaxation times. Observing ESR of the itinerant electrons is thus of great importance in graphene and in single-wall carbon nanotubes (SWCNTs). Often, the identification of CESR signal is based on two facts: the apparent asymmetry of the ESR signal (known as a Dysonian lineshape) and on the temperature independence of the ESR signal intensity. We argue that these are insufficient as benchmarks and instead the ESR signal intensity (when calibrated against an intensity reference) yields an accurate characterization. We detail the method to obtain the density of states from an ESR signal, which can be compared with theoretical estimates. We demonstrate the success of the method for K doped graphite powder. We give a benchmark for the observation of ESR in graphene.
\end{abstract}
\maketitle

\section{Introduction}
Electron spin resonance (ESR) has proven to be an important method in identifying the ground state of strongly correlated electron systems. ESR helped e.g. to identify the ordered spin-density wave ground state in the Bechgaard salts \cite{TorrancePRL1982} and for carbonaceous materials, ESR was key to discover the AC$_{60}$ (A=K, Rb, Cs) fulleride polymer \cite{JanossyPRL1994}.\\
\indent A natural expectation is that ESR can be applied for single-wall carbon nanotubes (SWCNTs) \cite{IijimaNAT1993} and graphene \cite{NovoselovSCI2004}, which are the two novel members of the carbon nanostructure family. The ESR literature on graphene is yet restricted to a single report \cite{ForroPSSB2009}. Although there exists larger literature on the SWCNTs, the situation is yet unclear. In general, the ESR signal on itinerant electrons yields a direct measurement of the spin-relaxation time (often called as spin-decoherence time), $T_1$, through the relation: $T_1=1/\gamma \Delta B$, where $\Delta B$ is the homogeneous ESR line-width and $\gamma/2 \pi=28.0\,\text{GHz/T}$ is the electron gyromagnetic ratio. $T_1$ is the central parameter which characterizes the usability of the materials for spintronics. This explains the motivation of the ESR studies on graphene and SWCNTs.\\
\indent One important question is whether the ESR signal of the itinerant (i.e. the conduction electrons) can be observed at all. It was argued on a theoretical basis \cite{DoraPRL2008} that it cannot be observed due to the Tomonaga-Luttinger liquid ground state of the metallic SWCNTs \cite{BockrathNAT,KatauraNAT2003,PichlerPRL2004}. It seemed that the only way to explore the local magnetism in SWCNTs is to spin label it either by means of $^{13}$C nuclei \cite{SingerPRL2005} or by an electron spin label \cite{SimonPRL2006}. The literature situation on the SWCNT ESR studies is conflicting, and it is reviewed herein without any judgement on validity. Petit \textit{et al.} \cite{PetitPRB1997} reported the observation of the ESR signal of itinerant electrons. Salvetat \textit{et al.} \cite{SalvetatPRB2005} reported that the ESR signal occuring around $g \approx 2$ is caused by defects in the SWCNTs. Likodimos \textit{et al.} \cite{likodimos} reported that a similar signal is related to the itinerant electrons with a possible antiferromagnetic order at low temperature. Corzilius \textit{et al.} \cite{CorziliusPSSB2008} reported the observation of the itinerant electron ESR in SWCNT samples prepared by chemical vapor deposition.

Often, the identification of the itinerant electron ESR signal is based on two facts: the asymmetry of the ESR lineshape (also known as a Dysonian) and the temperature independence of the ESR signal intensity. The Dysonian lineshape also occurs for localized spins (e.g. for paramagnetic impurities) which are embedded in a metal thus this property cannot be used for the above identification. This is discussed as Eq.\ 3.6 in the seminal paper of Feher and Kip as the "slowly diffusing magnetic dipole case" \cite{FeherKip}. The temperature independence of the ESR intensity could be observed for localized paramagnetic spins when they are embedded in a metal with increasing conductivity, $\sigma$ with decreasing temperature; then the microwave penetration depth $\lambda=\sqrt{\frac{2}{\mu_0 \omega \sigma}}$ (here $\mu_0$ is the permeability of the vacuum and $\omega$ is the frequency of the microwaves).

There has been remarkable progress in the quest for the intrinsic ESR signal in SWCNTs using samples made of nanotubes separated according to their metallicity \cite{ArnoldSeparation}. However, both kinds of samples, i.e. those made of purely metallic or semiconducting nanotubes shows similar ESR signals \cite{KuzmanyPSSB2010}, thus the situation remains unresolved.

A parallel situation happened for high $T_c$ superconductors: soon after their discovery \cite{BednorzMueller1986} several reports claimed to have observed the "intrinsic" ESR signal in these compounds. Later it turned out for all studies that the signal of parasitic phases (which happen to have strong paramagnetic signals), the so-called green and brown-phases were observed. Later, spin labeling (e.g. Gd substituting Y in YBa$_2$Cu$_3$O$_{7-\delta}$) turned out to be successful to study the electronic structure \cite{JanossyGdYBCO}.

The ESR signal of itinerant electrons in the SWCNTs is expected to have i) a $g$-factor near 2, ii) a line-width, $\Delta B$ smaller than 1 mT, and iii) a signal intensity corresponding to the low density of states (DOS) with no temperature dependence. All properties present a significant hindrance for the signal identification since most impurity in carbon have $g\approx 2$, a maximum 1 mT line-width, and the Curie spin-susceptibility of even a small amount of impurity overwhelms the small Pauli susceptibility of the itinerant electrons. Since nothing is known about the $g$-factor and the line-width \textit{a priori}, only the magnitude of the calibrated ESR signal when compared to the theoretical estimates of the Pauli spin-susceptibility provides a clear-cut ESR signal identification in graphene or SWCNTs.

Here, we outline the method to determine the calibrated ESR signal intensity and the resulting DOS in one- and two-dimensional carbon. The method is demonstrated for K doped graphite powder which is regarded as a model system of biased graphene \cite{GrueneisPRB2009}. A good agreement is obtained between the theoretical and expeirmental DOS for the KC$_8$ doped graphite system. We note, that a similar program was applied successfully when the ESR signals of Rb$_3$C$_{60}$ \cite{JanossyPRL1993} and MgB$_{2}$ \cite{SimonPRL2001} were discovered. We give benchmarks which can be used to decide whether the ESR of the itinerant electrons is observed in graphene.

\section{Experimental}
We used commercial graphite powder (Fischer Scientific) and potassium (99.95 \% purity: Sigma-Aldrich) for the intercalation experiments. The graphite powder (3 mg) was mixed with 3 mg MnO:MgO powder (Mn concentration 1.5 ppm) and ground in a mortar. MgO separates the graphite powder pieces, which enables the penetration of exciting microwave and its Mn content acts as an ESR intensity standard. The mixture was vacuum annealed at 500 $\mathrm{^{\circ}C}$ for 1 h in an ESR quartz tube and inserted into an Ar glove-box without air exposure. Alkali doping was performed by heating the ESR quartz tube containing the graphite powder and potassium for 29 hours using the standard temperature gradient method in Ref. \cite{Dresselhaus2002} to obtain Stage I, i.e. KC$_8$ intercalated graphite. ESR measurements were performed with a JEOL X-band spectrometer at room temperature.

\section{Results and discussion}
\begin{table*}

  \caption{\label{susceptibilities} The Curie and the Pauli spin-susceptibilities in three and two dimensions. Note that in two dimensions $A_c$ replaces $V_c$ in the expressions.}
  \begin{ruledtabular}
  \begin{tabular}{cccccccccccc}
   \noalign{\smallskip}
    \multicolumn{1}{c}{} \hspace{1.2cm}& \multicolumn{2}{c}{Curie susceptibility} & \multicolumn{2}{c}{} \hspace{2.2cm}& \multicolumn{2}{c}{Pauli susceptibility}& \multicolumn{2}{c}{}\hspace{2cm}&\multicolumn{2}{c}{Units}&\multicolumn{1}{c}{} \\
       \cline{2-3} \cline{6-7} \cline{10-11}
      \multicolumn{2}{c}{SI}&\multicolumn{2}{c}{Gaussian}&\multicolumn{2}{c}{SI}&\multicolumn{2}{c}{Gaussian}&\multicolumn{2}{c}{SI 3D (2D)}&\multicolumn{2}{c}{Gaussian 3D (2D)}\\
    \hline\noalign{\smallskip}
   \multicolumn{2}{c}{$\displaystyle \mu_0 \frac{g^2 S(S+1)\mu_{\text{B}}^2}{\displaystyle 3 k_{\text{B}}T}\frac{1}{V_c}$} & \multicolumn{2}{c}{$\displaystyle \frac{g^2 S(S+1)\mu_{\text{B}}^2}{\displaystyle 3 k_{\text{B}}T}\frac{1}{\displaystyle V_c}$}& \multicolumn{2}{c}{$\displaystyle \mu_0 \frac{g^2\mu_{\text{B}}^2}{4}\varrho(\varepsilon_{\text{F}})\frac{1}{\displaystyle V_c}$}& \multicolumn{2}{c}{$\displaystyle \frac{g^2\mu_{\text{B}}^2}{4}\varrho(\varepsilon_{\text{F}})\frac{1}{\displaystyle V_c}$} & \multicolumn{2}{c}{1(m)} &\multicolumn{2}{c}{$\mathrm{\displaystyle \frac{emu}{\displaystyle cm^3\cdot Oe}}\left(\mathrm{\displaystyle \frac{emu}{\displaystyle cm^2\cdot Oe}}\right)$}\\
    \noalign{\smallskip}
  \end{tabular}
  \end{ruledtabular}

\end{table*}

First, we discuss spin-susceptibility, $\chi_s$, calculated from the ESR signal in different dimensions. ESR spectroscopy measures the net amount of magnetic moments, which is an extensive thermodynamic variable, i.e. proportional to the sample amount. The corresponding intensive variable, which characterizes the material is the spin-susceptibility, $\chi_s$ which reads as:
\begin{equation}
\begin{split}
\chi_s=\mu_0\cdot\frac{\sum m}{B_{\text{res}}\cdot V_D} \hspace{1.5cm} \mathrm{(SI)}\\
\chi_s=\frac{\sum m}{B_{\text{res}} \cdot V_D} \hspace{1.5cm}\mathrm{(Gaussian)}
\end{split}
\end{equation}
\noindent where $m$ is the magnetic moment, $B_{res}$ is the magnetic field of the resonance, $V_D$ is the volume in $D$ dimension ($D=2;3$), and $\mu_0$ is the permeability of the vacuum. Clearly, the unit of $\chi_s$ depends on the dimension $D$.

$\chi_s$ is either due to the Curie spin-susceptibility for non-interacting spins or the Pauli spin-susceptibility for itinerant electrons in a metal. The relevant expressions are given in Table \ref{susceptibilities}. 
Therein, $A_c$/$V_c$ denotes the unit area/volume, $g$ is the $g$-factor, $\mu_{\text{B}}$ is the Bohr moment and $k_{\text{B}}$ is the Boltzmann constant. $S$ is the spin state of the non-interacting spins and $\varrho(\varepsilon_{\text{F}})$ is the DOS at the Fermi level in units of $\text{states} /\text{eV}\cdot unit$. Here, $unit$ refers to the unit chosen, e.g. for C$_{60}$ fulleride salts, the unit could be 60 carbon atoms. Then the DOS is larger but so is the unit volume which cancels in the result. For graphene, the two atom basis is used as $unit$.

The ESR intensity of a metal can be calibrated against a Curie spin system with known amount of spins. This leads to the comparison of the Pauli and the Curie spin-susceptibilities:

\begin{equation}
\begin{split}
\frac{I_{\text{ESR}}(\text{Pauli})}{I_{\text{ESR}}(\text{Curie})}=\frac{\sum m_{\displaystyle \text{Pauli}}}{\sum m_{\displaystyle \text{Curie}}}=\left(\frac{g_{\text{Pauli}}}{ g_{ \text{Curie}}}\right)^2\times\hspace{1.5cm}\\
\noalign{\smallskip}
\frac{4}{3}S(S+1)\cdot k_{\text{B}}T\varrho(\varepsilon_{\text{F}})\frac{B_{\mathrm{res}}(\text{Pauli})}{B_{\mathrm{res}}(\text{Curie})}\cdot
\!\frac{ \left(\frac{\displaystyle V_D}{\displaystyle V_c(D)}\right)(\text{Pauli})}{\displaystyle \left(\frac{\displaystyle V_D}{\displaystyle V_c(D)}\right)(\text{Curie})}
\label{chi_Curie_div_Pauli}
\end{split}
\end{equation}

\noindent where $I_{\text{ESR}}$ denotes the ESR signal. $V_D$ and $V_c(D)$ are the volume of the sample and the unit cell in $D$ dimensions, respectively. Note that $V_D/V_c(D)(\text{Pauli})=N(\text{ Pauli})$ is the number of units in the metallic sample and $V_D/V_c(D)(\text{Curie})=N(\text{Curie})$ is the number of Curie spins. Eq.\ (\ref{chi_Curie_div_Pauli}) is correct for both SI and Gaussian units and is independent of the choice of $unit$, as expected.

 For $S=1/2$ and $g_{\text{Pauli}},g_{\text{Curie}}\approx 2$, Eq.\ (\ref{chi_Curie_div_Pauli}) simplifies to:
\begin{equation}
\begin{split}
\frac{I_{\text{ESR}}(\text{Pauli})}{I_{\text{ESR}}(\text{Curie})}= k_{\text{B}}T
\varrho(\varepsilon_{\text{F}}) \frac{N(\text{ Pauli})}{N( \text{ Curie})}.
\label{chi_C_div_P_simple}
\end{split}
\end{equation}

\begin{figure}[t]%
\includegraphics*[width=\linewidth,height=0.7\linewidth]{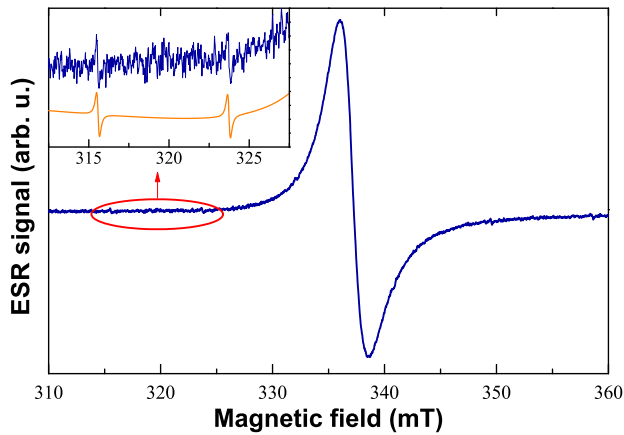}
\caption{\label{GraphDoped}
  ESR spectrum of saturated K doped graphite powder sample at $T=$300 K. The inset shows a zoom on the ESR spectrum showing the presence of the six lines of the $\mathrm{Mn^{2+}}$ hyperfine structure. The solid curve is a fit.}

\end{figure}
\begin{table}

  \caption[]{  \label{calib2} Parameters of the $\chi_s$ calibration of KC$_8$.}
  \begin{ruledtabular}
  \begin{tabular}{l|cc}
   \multicolumn{1}{l}{}&Mn:MgO&Graphite\\
   \hline\noalign{\smallskip}
      $\left\langle S\cdot(S+1)\right\rangle$& 9/4& \\
      $M_{\mathrm{mol}}$ [g/mol]& 40& 12\\
      $m$ [mg]&3&3\\
      $C_{\mathrm{spin}}$ & 1.5 ppm& 1\\
      $I_{\mathrm{ESR}}$& $6\cdot 5.3\cdot10^{-3}$&205\\

  \end{tabular}
  \end{ruledtabular}
\end{table}

\noindent We present the case of KC$_8$ as an example of the ESR intensity calibration. In Fig.\ \ref{GraphDoped}, we show the ESR signal of the mixture of MnO:MgO and saturated K doped graphite. Parameters of the calibration are given in Table \ref{calib2}: $C_{\mathrm{spin}}$ is the spin concentration and the effective $\left\langle S\cdot(S+1)\right\rangle_{\mathrm{Mn^{2+}}}=9/4$ as only the $-1/2 \rightarrow 1/2$ transition is observed from the 5 Zeeman transitions of the Mn$^{2+}$ ($S=5/2$) \cite{AbragamBleaneyBook}.

 The sample content gives: $N(\text{Pauli})/N( \text{Curie})\approx3.33$ and Eq.\ (\ref{chi_Curie_div_Pauli}) yields $\varrho(\varepsilon_{\text{F}})\approx0.34(2)\;\mathrm{states/(eV\cdot C\:atom)}$, in good agreement with  $\varrho(\varepsilon_{\text{F}})=0.327\;\mathrm{states/(eV\cdot C\:atom)}$ obtained by specific heat measurements \cite{Dresselhaus2002}.

In the following, we analyze the case of graphene. There, $A_c = 5.24$ \AA$^2/(\text{unit cell})$ is the graphene elementary cell and the DOS, at $T=0$ and $\Gamma=0$ ($\Gamma$ is the damping parameter), reads as a function of the chemical potential $\mu$ \cite{NovoselovRMP2009}:
\begin{equation}
\begin{split}
\varrho(\mu)=\frac{2A_c \mu}{\pi \hbar^2 v^2_{\text{F}}}.
\label{graphene_DOS}
\end{split}
\end{equation}
\noindent Here, $v_{\text{F}}\approx 10^{6}\,\text{m/s}$ is the Fermi velocity. Consequently, $\varrho(\mu)=\mu \cdot 0.0770\,\text{states/(eV}^2\cdot \text{unit cell})$ if $\mu$ is measured in eV. Thus Eq.\ (\ref{chi_C_div_P_simple}) (in two dimensions) at room temperature reads:

\begin{equation}
\begin{split}
\frac{\sum m_{\displaystyle \text{gr}}}{\sum m_{\displaystyle \text{Curie}}}=\underbrace{0.026 \cdot 0.0770}_{\approx 1/500}
\cdot \mu[\text{eV}] \frac{N(\text{gr})}{N(\text{Curie})}
\label{chi_graphene_div_Curie}
\end{split}
\end{equation}
\noindent $N(\text{gr})$ is the number of graphene unit cells in the sample.

Finally, we assess the feasibility of ESR spectroscopy on graphene. ESR spectrometer performance is given by the limit-of-detection (LOD$_0$) i.\ e.\ the number of $S=1/2$ Curie magnetic moments at room temperature which are required for a signal-to-noise ratio of $S/N =10$ for $\Delta B =0.1$ mT linewidth, and 1 s/spectrum-point time constant. For modern spectrometers LOD$_0=10^{10}$ spins/0.1 mT. To calculate the LOD for a broadened ESR line, LOD$(\Delta B)$, we introduce a function to track the effect of broadening:

\begin{equation}
\begin{split}
f(\Delta B)=\left\{
     \begin{array}{lr}
       \frac{\displaystyle \Delta B}{\displaystyle 0.1 \,\text{mT}} & \hspace{0.5 cm}\text{if}\hspace{0.5 cm} \Delta B\leq 1\,\text{mT}\\
       \noalign{\smallskip}
       \frac{\displaystyle \Delta B^2}{\displaystyle 0.1\,\text{mT}^2} & \hspace{0.5 cm}\text{if}\hspace{0.5cm} \Delta B> 1\,\text{mT}\\
     \end{array}
   \right.
\label{line_width_function}
\end{split}
\end{equation}

This function is 1 if $\Delta B=0.1\,\text{mT}$ and it is 10 if $\Delta B=1\,\text{mT}$ which is the usual maximum modulation amplitude. For line-widths above this value, the function grows quadratically, which describes that the amplitude of the derivative ESR signal drops quadratically. Using this function: LOD$(\Delta B)=$LOD$_0\cdot f(\Delta B)$. Comparison with Eq.\ (\ref{chi_graphene_div_Curie}) yields that numerically ($\mu$ in eV units)
\begin{equation}
\begin{split}
\text{LOD}(\text{gr})=500/\mu \cdot \text{LOD}_0 \cdot f(\Delta B)
\label{LOD_graphene}
\end{split}
\end{equation}
is the LOD for graphene. We could conclude that
\begin{equation}
\begin{split}
A_{\text{lb}}(\text{gr})=500/\mu \cdot \text{LOD}_0 \cdot f(\Delta B) \cdot A_c
\label{Alb_graphene}
\end{split}
\end{equation}
which gives a lower bound for the area of the graphene sheet which enables the ESR measurement. Assuming a $\Delta B=0.1$ mT and a shift in chemical potential by gate bias of $\sim$0.2 eV we estimate $A_{\text{lb}}(\text{gr})\approx1.3\,\mathrm{mm^2}$.

\section{Summary}
In summary, we detailed the method of obtaining the calibrated ESR intensity and the DOS in carbonaceous materials. We argue that a similar analysis is required for the identification of the ESR signal of itinerant electrons in SWCNT and graphene.

\section{Acknowledgements}
Work supported by the OTKA Grant Nr. K 81492, and Nr. K72613, by the ERC Grant Nr. ERC-259374-Sylo, the Marie Curie ERG project
CARBOTRON, and by the New Sz\'{e}chenyi Plan Nr. T\'{A}MOP-4.2.1/B-09/1/KMR-2010-0002. BD acknowledges the Bolyai programme of the Hungarian Academy of Sciences. The Swiss NSF and its NCCR "MaNEP" are acknowledged for support.

\end{document}